\documentclass[pra,aps,showpacs,groupedaddress,superscriptaddress,twocolumn,toc=flat,biblatex,footinbib]{revtex4-1}
\usepackage{natbib}
\usepackage[table]{xcolor}
\usepackage[colorlinks=false]{hyperref}
\usepackage[utf8]{inputenc}
\usepackage{color}
\usepackage{bbm} 
\usepackage[english]{babel}
\usepackage{graphicx}
\usepackage{placeins}
\usepackage{multirow}
\usepackage{csquotes}
\usepackage{physics}
\usepackage{appendix}

\usepackage{amsfonts,amsmath,amssymb,stmaryrd}

\usepackage{graphicx}
\usepackage{subfigure}  

\usepackage{epsfig}
\usepackage{mathrsfs}
\usepackage{verbatim}
\usepackage{centernot}
\usepackage{ulem}
\usepackage{longtable}
\usepackage{nicefrac}
\usepackage[framemethod=tikz]{mdframed}

\usepackage{tikz}

\usepackage{array}

\usepackage{cancel,ifthen}
\newcommand{\cmnt}[2][NoInPuT]{\ifthenelse{\equal{#1}{NoInPuT}}{}{{\color{red}\sout{#1}}} {\color{blue} #2}}

\usepackage{bm}	
\renewcommand{\vec}[1]{\bm{#1}}

\LTcapwidth=\textwidth

\begin{document}

\normalem	

\title{Pairing dome from an emergent Feshbach resonance \\ in a strongly repulsive bilayer model}

\author{Hannah Lange}
\affiliation{Ludwig-Maximilians-University Munich, Theresienstr. 37, Munich D-80333, Germany}
\affiliation{Max-Planck-Institute for Quantum Optics, Hans-Kopfermann-Str.1, Garching D-85748, Germany}
\affiliation{Munich Center for Quantum Science and Technology, Schellingstr. 4, Munich D-80799, Germany}

\author{Lukas Homeier}
\affiliation{Ludwig-Maximilians-University Munich, Theresienstr. 37, Munich D-80333, Germany}
\affiliation{Munich Center for Quantum Science and Technology, Schellingstr. 4, Munich D-80799, Germany}

\author{Eugene Demler}
\affiliation{Institute for Theoretical Physics, ETH Zurich, 8093 Zürich, Switzerland}

\author{Ulrich~Schollwöck}
\affiliation{Ludwig-Maximilians-University Munich, Theresienstr. 37, Munich D-80333, Germany}
\affiliation{Munich Center for Quantum Science and Technology, Schellingstr. 4, Munich D-80799, Germany}

\author{Annabelle Bohrdt}
\affiliation{Munich Center for Quantum Science and Technology, Schellingstr. 4, Munich D-80799, Germany}
\affiliation{University of Regensburg, Universitätsstr. 31, Regensburg D-93053, Germany}

\author{Fabian Grusdt}
\affiliation{Ludwig-Maximilians-University Munich, Theresienstr. 37, Munich D-80333, Germany}
\affiliation{Munich Center for Quantum Science and Technology, Schellingstr. 4, Munich D-80799, Germany}


\date{\today}

\begin{abstract}
A key to understanding unconventional superconductivity lies in unraveling the pairing mechanism of mobile charge carriers in doped antiferromagnets, yielding an effective attraction between charges even in the presence of strong repulsive Coulomb interactions. Here, we study pairing in a mixed-dimensional (mixD) $t-J$ model, featuring robust binding energies -- despite dominant repulsive interactions -- that are strongly enhanced in the finite doping regime. The single and coupled mixD ladders we study, corresponding to bilayers of width $w\leq 2$, feature a crossover from tightly bound pairs of holes (closed channel) at small repulsion, to more spatially extended, correlated pairs of individual holes (open channel) at large repulsion. We derive an effective model for the latter, in which the attraction is mediated by the closed channel, in analogy to atomic Feshbach resonances. Using density matrix renormalization group (DMRG) simulations we reveal a dome of large binding energies at around $30\%$ doping, accompanied by a change of the Fermi surface volume and a crossover from extended to tightly bound hole pairs. Our work provides a microscopic theory of pairing in the doped mixD system with dominant repulsion, closely related to bilayer, Ni-based superconductors, and our predictions can be tested in state-of-the-art quantum simulators.  
\end{abstract}

\maketitle


\textbf{\textit{Introduction.--}} Among the remaining mysteries of high-T$_c$ superconductivity \cite{Bednorz1986,Lee2006,Scalapino1999} is the pairing mechanism of charge carriers, leading to the formation of Cooper pairs \cite{Cooper1956} in a relatively high temperature regime and despite repulsive Coulomb interactions between the charges \cite{Kohn1965, Kantian2019, Chakravarty2001}. In contrast to conventional superconductors, for which BCS theory \cite{Bardeen1955} predicts small pairing gaps that result in large Cooper pairs -- effectively circumventing long-range Coulomb repulsion --, high-T$_c$ superconductors are characterized by their substantial pairing gap \cite{Hashimoto:014} and Cooper pairs are potentially exposed to extended-range Coulomb interactions. To investigate pairing mechanisms resilient to such Coulomb interactions, Fermi Hubbard or $t-J$ type models have been extended to $t-J-V$ models \cite{Feiner} with nearest-neighbor repulsion $V$. These models have been shown to sustain pairing up to large repulsion strengths \cite{Wang2001,Zinni2021} and have proven to effectively describe some experimental results in cuprates, e.g. on plasmon spectra \cite{Hepting2023,Greco2016,Greco2019}. However, such models are prohibitively difficult to solve in order to unravel the underlying pairing mechanism, despite impressive numerical advances in the past years \cite{Qin2020,Schaefer2021,xu2023,Arovas2022}. 

\begin{figure}[t]
\centering
\includegraphics[width=0.45\textwidth]{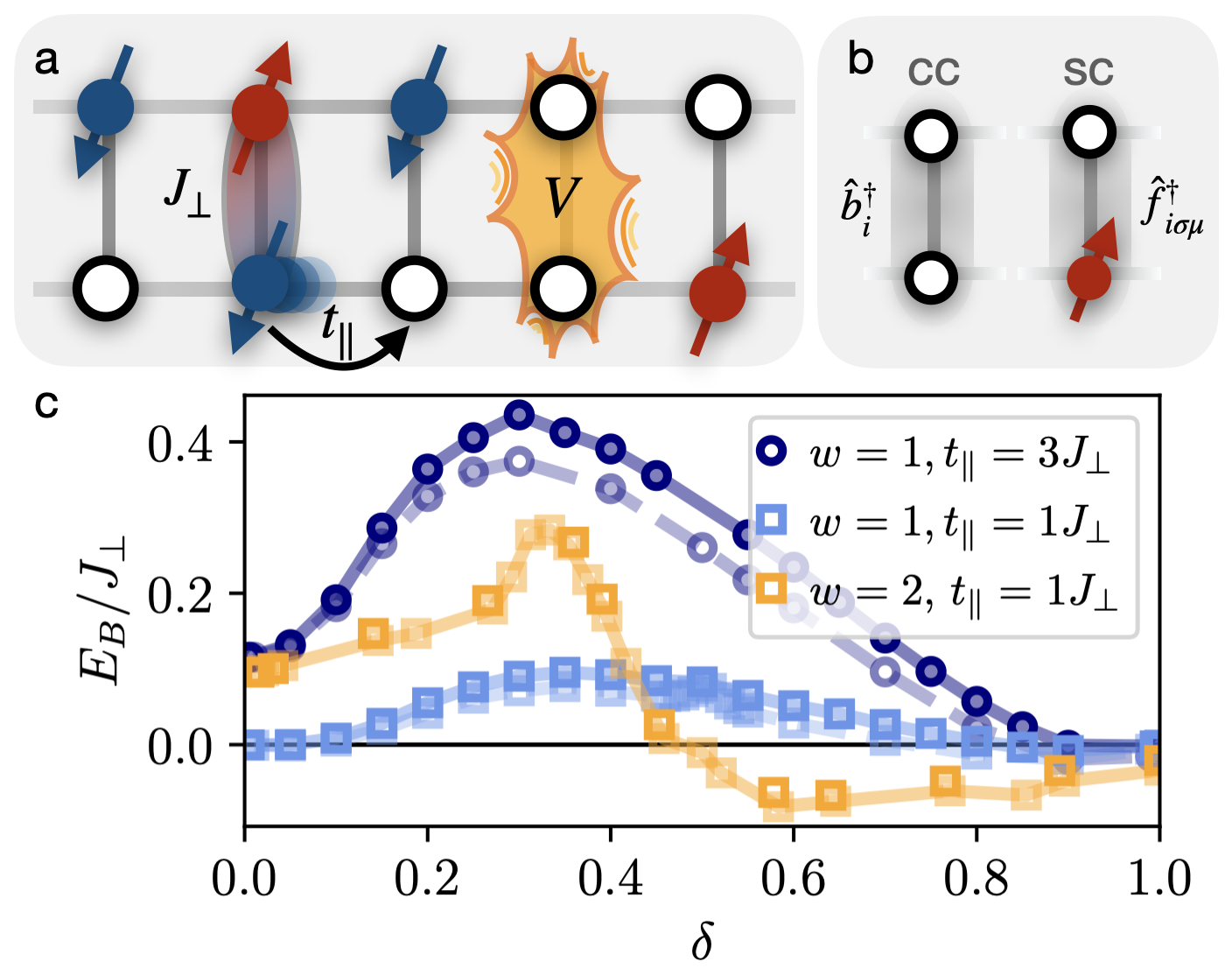}
\vspace{-0.4cm}
\caption{a) Schematic picture of the mixD ladder, with exchange coupling $J_\perp$ between the legs, hopping $t_\parallel$ only along the legs and repulsion $V$ between holes on the same rung. b) The constituents of the effective theory in the large $V$ limit: chargon-chargon (cc) pairs $\hat{b}_i^{(\dagger)}$ and spinon-chargon (sc) pairs $\hat{f}_{\sigma\mu i}^{(\dagger)}$. c) Binding energies w.r.t. hole doping $\delta$ for $V/J_\perp=5$, for ladders (blue) with $L_x=100 (200)$ and bilayers of width $w=2$ (orange) with $L_x=24(32)$ (dashed (solid) lines). } 
\vspace{-0.6cm}
\label{fig:1}    
\end{figure}
Here, we study a pairing mechanism that gives rise to attractively interacting fermions despite dominant repulsion between them, with enhanced pairing in the intermediate doping regime, leading to the dome of binding energies in Fig. \ref{fig:1}. For this purpose, we consider single and coupled ladders at high repulsion. Despite their simplicity, $t-J$ and Fermi-Hubbard ladders have been shown to feature a variety of correlated phases \cite{Dagotto1992,Sigrist1994,Zhu2014,Zhu2022,Zhang2022,Jiang2020}, including superconducting correlations even when supplementing the models with repulsive Coulomb interactions between nearest neighbors \cite{Dagotto1992_2,Troyer1194}. Moreover, so-called mixed-dimensional (mixD) ladders, featuring hopping only along the legs but magnetic superexchange in both directions, have been shown to host binding on extremely high energy scales, even exceeding the superexchange energy \cite{Grusdt2018_2,Chen2018,Zhu2018,Bohrdt2021, Bohrdt2022,Hirthe2023,JIANG2018753}. This allows to observe binding \cite{Bohrdt2022, Hirthe2023} and stripe formation \cite{Schloemer2022,bourgund2023formation} in these models using ultracold atom experiments \cite{Hart:2015,Mazurenko:2017aa,Bloch2008,Gross2017,Bohrdt2021}, where experimentally reachable temperatures are today typically on the order of the superexchange energy \cite{Mazurenko:2017aa}. 
Furthermore, recent works proposed that the high-temperature superconductor La$_3$Ni$_2$O$_7$ \cite{Sun2023,zhang2023hightemperature} can be modeled by mixD bilayers, see e.g. Refs. \cite{Sun2023,Wu2023,Lu2023,Qu2023,Luo2023,Gu2023,yang2023strong,schloemer2023superconductivity}, and might hence be closely related to the mixD bilayers of width $w=1,2$ studied in this work.

Here, we utilize the mixD model as a controlled setting to study the rich interplay of magnetic fluctuations, Coulomb repulsion and arbitrary doping, see Fig.~\ref{fig:1}, with interesting emergent structures: When the half-filled ground state of a ladder with small intra-leg superexchange $0\leq J_\parallel\ll J_\perp$ is doped with a single hole, the latter can be understood as a bound state of two partons, a \textit{chargon} and \textit{spinon}, carrying the respective quantum numbers and connected by a linear confinement potential \cite{Bohrdt2021,Bohrdt2022}, as observed also in 2D \cite{Brinkman1970,Trugman1988,Beran1996, Laughlin1997,Senthil2003, Grusdt2018, Grusdt2019,Chiu2019}. Similarly, two holes form a pair of two chargons in the mixD setting \cite{Chen2018,Zhu2018,JIANG2018753,Bohrdt2022,Hirthe2023}. In analogy to mesons in high energy physics, the constituents (\textit{partons}) of this pair are very tightly bound, and hence the chargon-chargon bound state will also be referred to as \textit{meson} in the following. In this letter we show that strong interactions -- approximating the Coulomb repulsion of electrons in solids -- favor another pairing scenario which, we argue, is more closely related to the pairing seen in most real materials: more extended Cooper-type bound states of two mesons, i.e. consisting of four constituents (\textit{tetrapartons}). 

Specifically, we investigate mixD $t-J$ bilayers of width $w=1,2$, i.e. single and coupled ladders, supplemented with strong repulsive interactions $V$ between two holes on a rung (see Fig.~\ref{fig:1}a) -- i.e. an off-site repulsion that may also be relevant in nickelate superconductors. Combining DMRG simulations \cite{White1992,Schollwoeck2011}, including the calculation of angle resolved photoemission spetroscopy (ARPES) spectra, and effective descriptions in terms of the emergent charge carriers we show that the effective attraction between the holes in the large $V$ regime is induced by coupling processes to the chargon-chargon, meson-like states, leading to a dome of binding energies in the intermediate doping regime. To highlight the analogy of this mechanism to Feshbach resonances \cite{Feshbach1958}, we refer to the high energy channel of chargon-chargon (cc) states as closed, mesonic channel and to the low energy spinon-chargon (sc) pairs as the open, tetraparton channel at large $V$. 

Overall, despite their simplicity, the mixD systems we study show some remarkable phenomenological similarities with strongly correlated superconductors such as cuprates and nickelates: \textit{(i)} We report pairing facilitated by doping, leading to a dome of binding energies with its peak at $30\%$ doping (see Fig \ref{fig:1}c);  \textit{(ii)} the pairing we find only requires short-range antiferromagnetic (AFM) correlations but no long-range magnetic order; \textit{(iii)} we observe a change of the Fermi surface volume, similar to the small-to-large Fermi surface transition in cuprates; (iv) we discover a density wave (with bond order) in the intermediate doping regime \cite{lange2023feshbach}. \\

\textbf{\textit{Model and emergent constituents.--}} The primary goal of this work is to investigate the impact of strong repulsive interactions $V$ on the pairing mechanism in a mixD system, featuring inter-layer exchange interactions but no inter-layer tunneling \cite{Bohrdt2022,Hirthe2023,Grusdt2018_2}, see Fig.~\ref{fig:1}a,
\begin{align}
    \hat{\mathcal{H}}&=-t_{\parallel}\hat{\mathcal{P}}\sum_{\langle ij\rangle }\sum_{\mu, \sigma}\left( \hat{c}_{i\mu\sigma}^\dagger \hat{c}_{j\mu\sigma} + \mathrm{h.c.}\right)\hat{\mathcal{P}}\notag \\
    &+ J_{\perp}\sum_{j }\left(\hat{\vec{S}}_{j0}\cdot \hat{\vec{S}}_{j1} -\frac{1}{4}\hat{n}_{j0}\hat{n}_{j1}\right) +V \sum_{j }\hat{n}_{j0}^h \hat{n}_{j1}^h\, .
        \label{eq:mixD_Hamiltonian}
\end{align}
Here, $\hat{\mathcal{P}}$ is the Gutzwiller projector onto the subspace with maximum single occupancy per site. Spin operators at site $j$ in layer $\mu=0,1$ are denoted by $\hat{\vec{S}}_{j\mu}$, (hole) density operators by $\hat{n}_{j\mu}=\hat{n}_{j\mu\uparrow}+\hat{n}_{j\mu\downarrow}$ ($\hat{n}_{j\mu}^h=1-\hat{n}_{j\mu}$), and $\hat{c}_{j\mu\sigma}$ annihilate a fermion with spin $\sigma=\uparrow,\downarrow$. 

At half-filling, the ground state of the system is given by spin singlets on each rung \cite{Bohrdt2022}. At finite doping, the system is dominated by a competition of kinetic energy and the energy of the distortion of the spin background when holes move. The emergent constituents in this doping regime are most easily understood when $t_\parallel\ll J_\perp$. At $V=0\leq t_\parallel\ll J_\perp$ two holes tend to sit on the same rung -- a configuration with energy $-J_\perp$ -- and form a chargon-chargon pair (cc), see Fig.~\ref{fig:1}b left. They can move freely through the system, since the second chargon restores the spin-singlet background when following the first one, making it favorable for charges to move through the system together, i.e. yielding large binding energies \cite{Bohrdt2022}. When the repulsive interaction $V$ reaches a critical value $V_c$, it is energetically favorable to place at maximum one hole (and one spin) per rung, i.e. to form a spinon-chargon pair (sc), see Fig.~\ref{fig:1}b right. In contrast to cc's, the motion of sc's is suppressed by the distortion of the singlet spin-background created when the chargon moves. 

When $t_\parallel>J_\perp$, this meson picture remains qualitatively correct, although the cc's and sc's develop a finite spatial extent determined by the interplay of the kinetic energy and a linear confining potential (a \textit{string}) between the partons \cite{Bohrdt2022,Grusdt2019}; see also \cite{lange2023feshbach}. As we show in the following sections, the crossover from the cc ($V<V_c$) to the sc regime ($V>V_c$) depends strongly on the ratio $t_\parallel/J_\perp$, i.e. $V_c = V_c\left( t_\parallel/J_\perp\right)$. \\

\begin{figure}[t]
\centering
\includegraphics[width=0.5\textwidth]{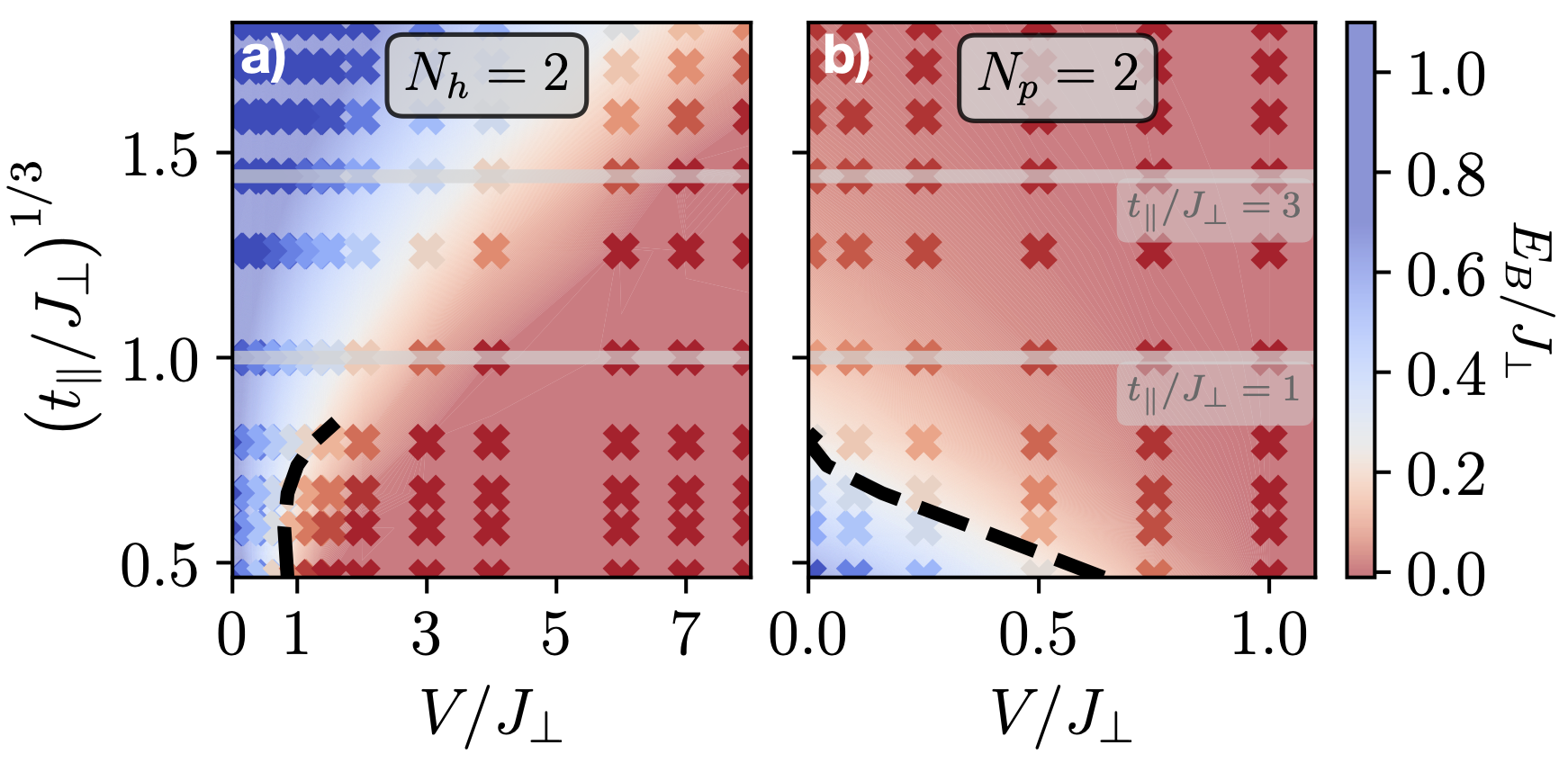}
\vspace{-0.7cm}
\caption{Binding energy in the limits of a) low ($N_h=2$) and b) high ($N_p=L_x\cdot L_y-N_h=2$) doping. If $t_\parallel\ll V, J_\perp $ the critical $V_c$ values for which cc's do not constitute the ground state any more are denoted by the black dotted lines (see \cite{SM}). Here, we expect the crossover to the sc regime.} 
\label{fig:3}  
\vspace{-0.4cm}
\end{figure}

\textbf{\textit{Limits of low and high doping.-- }}  Using the DMRG package SyTen \cite{syten1, syten2} with $U(1)_{N_{\mu=1,2}}$ symmetry on each layer and global $U(1)_{S_z^\mathrm{tot}}$ symmetry, we calculate the binding energies,
\begin{align}
    E_B(N_h)=&2\left(E_{N_h-1}-E_{N_h-2}\right) -\left(E_{N_h}-E_{N_h-2}\right)\, ,
\end{align}
where $N_h$ is the number of holes doped in the system and $E_B>0$ indicates binding of holes. Details on the implementation can be found in \cite{lange2023feshbach}. 

Our results
in the limits of very low ($N_h=2$) and high ($N_p=L_x\cdot L_y-N_h=2$) doping are shown in Fig.~\ref{fig:3}. For $t_\parallel=0$ the cc at low doping and the particle pair at high doping are bound by the fact that it is energetically favorable to form singlets on the rungs. In both cases, the pairs unbind when $V\geq V_c(t_\parallel=0)=J_\perp$. 

For $t_\parallel\ll J_\perp$ we perform a Schrieffer-Wolff transformation, as detailed further below, to estimate the critical value $V_c$, where the nature of the constituents changes from tightly bound cc's to weakly bound sc's in the low and high hole doping regimes \cite{Bohrdt2021}, see \cite{SM}. The resulting perturbative expressions for $V_c$, indicated by the dashed black lines in Fig.~\ref{fig:3}, agree well with the numerical results for $t_\parallel \leq J_\perp$. At low doping, Fig. \ref{fig:3}a, even the backbending observed numerically is captured correctly. Furthermore, analytics and numerics show that the low and high doping regimes are distinctly different: $(a)$ For two holes, the string-based pairing mechanism \cite{Bohrdt2022} yields pronounced binding even for large $t_\parallel/J_\perp$. For example, for $t_\parallel/J_\perp=3$, a repulsion $V>7J_\perp$ is needed to suppress the binding energy below $E_B/J_\perp=10\,\%$, coinciding with the numerical observation of widely spread hole pairs over more than $20$ sites for a system of length $L_x=80$ (see \cite{lange2023feshbach}). $(b)$ In the high doping limit with only two particles remaining in the system, the binding energy is suppressed to very small values $E_B/J_\perp<0.01$ as soon as $V \lessapprox J_\perp$. For large $t_\parallel\geq J_\perp$, the perturbative description breaks down. \\

\textbf{\textit{Strong pairing at finite doping.--}}For large $V$, we show in Figs.~\ref{fig:1}c and \ref{fig:3} that neither very low nor very high doping permits significant pairing of charges. Remarkably, we find drastically enhanced binding energies at intermediate doping values in single and coupled ladders in Fig.~\ref{fig:1}c, where $V=5J_\perp$: In all cases, $E_B$ shows a pronounced dome around $\delta_\mathrm{opt}\approx 30\%$ doping, before decreasing again down to a vanishingly small or even negative value for the two-particle system. For ladders, binding is increased from a vanishingly small value (substantial value) for $t_\parallel/J_\perp=1$ ($t_\parallel/J_\perp=3$) by a factor $>10$ ($3$) in the finite doping regime. The same behavior can be observed for bilayers of width $w=2$, but with larger binding energies in the low doping regime. This can be understood by the fact that in 2D, the chargon moves on a Bethe lattice, renormalizing the hopping to $t\to \sqrt{2}t\approx 1.7t$ \cite{Grusdt2018} and stabilizing the binding as explained above.  \\

To gain a better understanding of the binding energies in the finite doping regime we derive an effective model for $V>t_\parallel,J_\perp$ -- a regime for which sc configurations are dominant. In this limit of large $V>J_\perp \gg t_\parallel$ we identify two subspaces of $(i)$ sc constituents at low energy and $(ii)$ cc configurations that are gapped by $\Delta E=V-J_\perp $. We perform a Schrieffer-Wolff transformation \cite{Schrieffer1966} of Eq.~\eqref{eq:mixD_Hamiltonian} assuming small $\vert t_\parallel/\Delta E\vert$ and integrate out the cc states, see SM \cite{SM}. We express the effective Hamiltonian in terms of sc operators $\hat{f}_{i\mu\sigma}^{(\dagger)}$, acting on the vacuum state consisting of rung-singlets. At $V>J_\perp \gg t_\parallel$ at maximum one sc is allowed per rung, which we enforce by the projector $\mathcal{\hat{P}}_f$ on the corresponding subspace. Moreover, the effective Hamiltonian only applies below $\delta \leq 50 \%$ doping, before cc's naturally proliferate. 
We find (see \cite{lange2023feshbach}):
\begin{align}           
\mathcal{\hat{H}}^{sc}_\mathrm{eff}&= \frac{t_\parallel}{2}\sum_{j} \sum_{\sigma,\mu} \mathcal{\hat{P}}_f\left(\hat{f}^\dagger_{j+1\mu\sigma}\hat{f}_{j\mu\sigma}+\mathrm{h.c.}\right)\mathcal{\hat{P}}_f \notag \\
&+\epsilon_0\sum_{j\mu} \hat{n}^f_{i\mu}
+\frac{t_\parallel^2}{J_\perp}\frac{3}{2}\sum_{j}\sum_{\mu \mu^\prime}\hat{n}^f_{j+1\mu}\hat{n}^f_{j\mu^\prime}
\notag \\
& -4t_\parallel^2\sum_{j}\left(
-\hat{\vec{J}}_{j+1}\cdot \hat{\vec{J}}_j+\frac{1}{4}
\right)\left[\frac{\mathcal{\hat{P}}^S_j}{V-J_\perp }+\frac{\mathcal{\hat{P}}^T_j}{V }\right].
\label{eq:H_eff_sc}
\end{align}
Here, we have defined
$\epsilon_0= J_\perp -\frac{t_\parallel^2}{J_\perp}\frac{3}{2}$ and the singlet and triplet projectors $\mathcal{\hat{P}}^S_j= -\hat{\vec{S}}_{j+1}\cdot \hat{\vec{S}}_{j}+\frac{1}{4}\hat{n}^f_{j+1}\hat{n}^f_{j}$ and $\mathcal{\hat{P}}^T_j= \hat{\vec{S}}_{j+1}\cdot \hat{\vec{S}}_{j}+\frac{3}{4}\hat{n}^f_{j+1}\hat{n}^f_{j}$,
with the sc density operators $\hat{n}^f_{j\mu}=\sum_\sigma \hat{f}_{j\mu\sigma}^\dagger \hat{f}_{j\mu\sigma}$, the
sc spin operators $\hat{\vec{S}}_{j}=\frac{1}{2}\sum_{\mu}\sum_{\sigma\sigma^\prime}\hat{f}_{j\mu\sigma}^\dagger\vec{\sigma}_{\sigma\sigma^\prime}\hat{f}_{j\mu\sigma^\prime}$ and isospin leg operators
\begin{align}
    \hat{\vec{J}}_{j}=\frac{1}{2}\sum_\sigma \sum_{\mu\mu^\prime}\hat{f}_{j\mu\sigma}^\dagger\vec{\sigma}_{\mu\mu^\prime}\hat{f}_{j\mu^\prime\sigma}\, .
    \label{eq:J_operators}
\end{align}

Eq.~\eqref{eq:H_eff_sc} describes hard-core, fermionic sc's, experiencing repulsive (3rd term) and attractive interactions $\propto \frac{t_\parallel^2}{V-J_\perp}$ and $\propto \frac{t_\parallel^2}{V}$ (4th term with negative sign for spin singlet / triplet configurations and leg singlets) that compete with each other. We emphasize that the attraction is mediated by virtual processes involving the high-energy cc channel shown in SM, Fig.~\ref{fig:derivation}a, similar to the attraction induced at a Feshbach resonance. This is also apparent from the term $\propto \vec{\hat{J}}_{j+1}\cdot \vec{\hat{J}}_j$ which penalizes neighboring sc's occupying the same leg, since only sc's from opposite legs can lower their energy by recombining virtually into the cc channel. Furthermore, a resonance occurs at $V\to J_\perp$, where the attractive interaction diverges and becomes dominant over the repulsion.

The Feshbach-mediated pairing mechanism in Eq. \eqref{eq:H_eff_sc} is in qualitative agreement with the doping dependence of the binding energies of the mixD bilayers with $w=1,2$ in Fig. \ref{fig:1}c, and is in principle even valid for the full 2D limit $w\to \infty$. Near the resonance, when the attraction is dominant, the attractive interactions predicted by Eq. \eqref{eq:H_eff_sc} are effectively enhanced when the number of holes in the system is increased, since the kinetic energy per hole decreases with doping (Pauli pressure). This suggests an increasing binding energy with doping, i.e. when the number of sc's increases, similar to the dome of $E_B$ in Fig. \ref{fig:1}c. The optimal doping $\delta_\mathrm{opt}$, corresponding to maximum binding energy, is reached when sc's begin to overlap spatially: In the effective model \eqref{eq:H_eff_sc} of point-like sc's, this suggests a maximum at $\delta_\mathrm{opt}=50\%$, in agreement with our numerical results on the full mixD system as well as on the effective sc model for $t_\parallel\ll J_\perp$ \cite{lange2023feshbach}. For larger $t_\parallel \geq J_\perp$, we expect that the mechanism remains essentially the same, but with two main changes: $(i)$ Since binding is stabilized by $t_\parallel$, the resonance shifts to higher values $V_c>J_\perp$, yielding positive binding energies even for large $V= 5J_\perp$ at finite doping that are on the same order of magnitude as in Fig. \ref{fig:1}, see \cite{lange2023feshbach}. $(ii)$ sc's extend over several rungs and $\delta_\mathrm{opt}$ shifts to smaller values $< 50\%$, see Fig. \ref{fig:1}c. \\

Another remarkable feature of the effective model \eqref{eq:H_eff_sc} is the emergent isospin $SU(2)$ symmetry. We find numerical indications that this $SU(2)$ symmetry is only approximately present in the full mixD system \eqref{eq:mixD_Hamiltonian}, see \cite{SM}, with a strong doping dependence, see also \cite{lange2023feshbach}. When higher orders in $t_\parallel/\Delta E$ are considered, the $SU(2)$ isospin symmetry of Eq. \eqref{eq:H_eff_sc} breaks down. \\

\begin{figure}[t]
\centering
\includegraphics[width=0.46\textwidth]{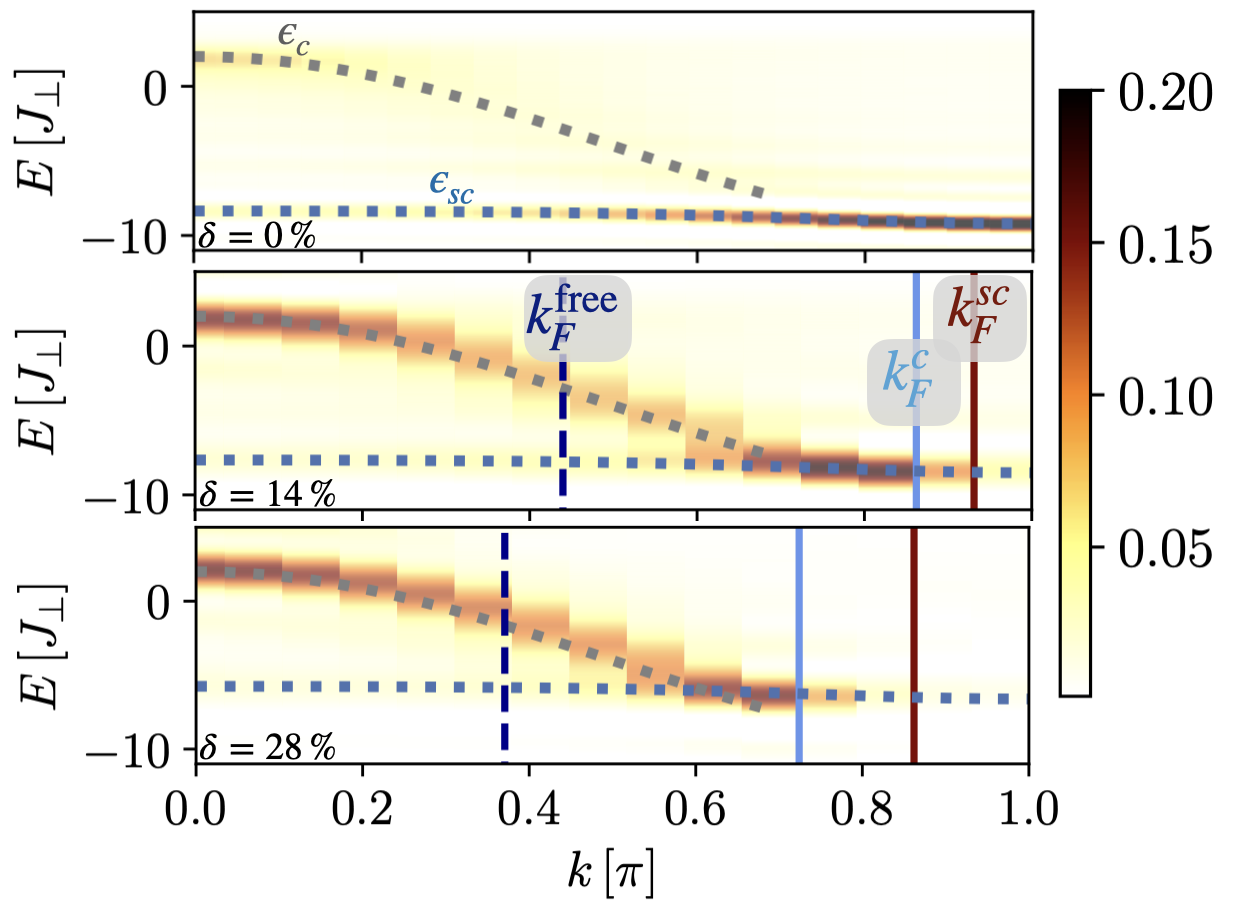}
\caption{ARPES spectra at $\delta=0, 0.14,0.28$ (top to bottom) and for $V/J_\perp=5$, $t_\parallel/J_\perp=3$. Furthermore, we show eigenenergies of free sc's, $\epsilon_c$, and sc's, $\epsilon_{sc}$, from geometric string theory \cite{Bohrdt2022} (dotted lines) as well as Fermi momenta for free fermions $k_F^\mathrm{free}$, free chargons $k_F^\mathrm{c}$ and sc pairs $k_F^\mathrm{sc}$. Both $k_F^\mathrm{c}$ and $k_F^\mathrm{sc}$ coincident with an onset of spectral weight in the finite doping regime.} 
\label{fig:arpes}
\end{figure}

\textbf{\textit{ARPES spectra.--}} In Fig. \ref{fig:arpes} the ARPES spectra upon removing a single hole from the mixD ladder at $\delta=0$ (top), $\delta=0.14$ (middle) and $\delta=0.28$ (bottom) hole doping, focusing on high repulsion $V/J_\perp=5$, are shown (see SM \ref{appendix:ARPES}). At low doping, the low-energy spectrum is dominated by a narrow band, matching the dispersion of sc's $\epsilon_{sc}$ calculated from geometric string theory as in \cite{Bohrdt2022}. For larger $\delta$, the spectral weight continuously shifts into a more dispersive, higher energy branch that matches the dispersion of free particles, $\epsilon_c=2t\mathrm{cos}(k)$. Furthermore, the onset of the spectral weight, indicating the Fermi momentum of the system, changes from a small momentum to a large one with doping: At low doping, the onset matches the sc Fermi momentum $k_F^{sc}=\frac{1}{4} \pi n_{sc}$ with $n_{sc} = N_h/L_x$, where the factor $1/4$ accounts for layer and spin indices of sc's. At larger $\delta$, a second onset of spectral weight at $k_F^c=\pi n_c$ becomes apparent, which can be associated with free, spinless chargons with density $n_c=\delta$. Hence, our results indicate a change of the Fermi surface volume around $\delta_\mathrm{opt}$ \cite{yang2023strong}, similar to the phenomenology of cuprates \cite{Chowdhury}. The momentum of free, spinful fermions $k_F^{\mathrm{free}}=\frac{1}{2}\pi(1-\delta)$ does not match the onset of the spectral weight in any case.  \\

\begin{figure}[t]
\centering
\includegraphics[width=0.48\textwidth]{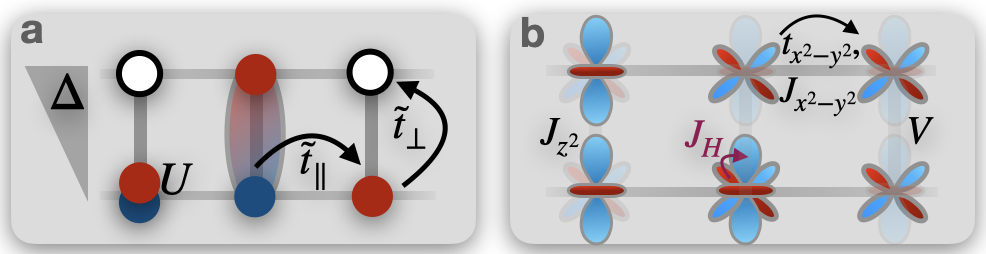}
\caption{ a) The ultracold atom setup for a mixD ladder with repulsion: a potential offset $\Delta$ and hole (doublon) doping in the upper (lower) leg are applied. b) Schematic illustration of the bilayer La$_3$Ni$_2$O$_7$: $d_{x^2-y^2}$ orbitals contribute to intralayer hopping and AFM exchange, $d_z^2$ orbitals to an interlayer AFM exchange and both orbitals are coupled by FM Hund's coupling $J_H$. } 
\label{fig:5}
\end{figure}

\textbf{\textit{Experimental Realizations.--}} Our studies are motivated by recent experiments in cold atoms and nickelate compounds: 

The mixD ladder without repulsion $V$ was already realized in a setup of ultracold fermionic atoms in an optical potential \cite{Hirthe2023} by applying a potential offset $\Delta$ between the legs to suppress the inter-leg tunneling $\Tilde{t}_\perp$ to an effective $t_\perp\approx 0$. To supplement this setup with a nearest-neighbors repulsion, we propose hole (doublon) doping for upper (lower) legs of the ladder, see Fig. \ref{fig:5}a. This gives rise to virtual hopping processes between doublons in the lower leg and holes in the upper leg with amplitude $2\frac{\Tilde{t}_\perp^2}{\Delta}$ and doublons and spins with $\frac{\Tilde{t}_\perp^2}{\Delta+U}$, yielding a total interaction strength 
$V=2 \Tilde{t}_\perp^2 \frac{U}{\Delta (U-\Delta)}$ \cite{lange2023feshbach}.

Furthermore, as discussed in \cite{Lu2023}, the recently discovered nickelate superconductor La$_3$Ni$_2$O$_7$ \cite{Sun2023} can be modeled by a $t_\parallel-J_\parallel-J_\perp$ bilayer similar to the model we study. In this material, the $d_{x^2-y^2} $
 orbitals form an effective intralayer $t_{x^2-y^2}-J_{x^2-y^2}$ model, whereas the $d_z^2$ orbitals are localized with interlayer antiferromagnetic (AFM) superexchange, see Fig. \ref{fig:5}b. Both orbitals interact via ferromagnetic (FM) Hund's coupling $J_H$. In the limit of sizable $\vert J_H \vert $ the spins of $d_{x^2-y^2} $ and $d_{z^2} $ form triplets, giving rise to an effective AFM interaction $J_\perp$ of $d_{x^2-y^2} $ spins between the layers \cite{Qu2023,Lu2023}. In contrast to AFM interactions solely originating from superexchange, the interaction mediated via Hund's rule yields vanishingly small interlayer hopping. We argue that in such a 2D material the Coulomb repulsion $V$ may play an important role at low doping. Our numerical results for increasing width $w$ of the bilayers and the effective model \eqref{eq:H_eff_sc}, which is also valid in the 2D limit $w\to \infty$, suggest that in such a 2D bilayer system, Feshbach-mediated pairing is a possible scenario \cite{Homeier2023}. In the BEC regime, similar effective models to Eq. \eqref{eq:H_eff_sc} have proven to provide an insightful perspective into the underlying physics in two dimensions \cite{schloemer2023superconductivity}. \\

\textbf{\textit{Summary and Outlook.--}}Our results show that effective attractive interactions between charge carriers can arise even in the presence of strong repulsive interactions, here in the setting of single and coupled mixD $t-J$ ladders, corresponding to bilayers of width $w=1,2$. The binding energies we obtain feature a pronounced maximum (pairing dome) in the intermediate doping regime at strong repulsion, where the system can be understood in terms of correlated sc's ($(\mathrm{sc})^2$). 

The state that we observe is a Luther-Emery liquid of $(\mathrm{sc})^2$ pairs with no charge gap but a spin gap, except for a bond-ordered density wave at $\delta=50\%$, which we discuss in \cite{lange2023feshbach}. Similar to works on atomic BEC-BCS crossovers (e.g. \cite{Fuchs2004,Recati2005,Tokatly2004,Citro2005}), where binding in the Luther-Emery state is induced by a narrow Feshbach resonance with a closed channel of bosonic molecules \cite{Citro2005}, our mixD model results suggest that binding in the $(\mathrm{sc})^2$ regime arises from coupling to the closed cc channel. Furthermore, our spectroscopic analysis shows that a change from a small to large Fermi surface occurs in the intermediate doping regime, which can be associated with sc's and free chargons, respectively.

Finally, our results may have implications for understanding pairing in high-$T_c$ cuprate compounds \cite{Homeier2023} and recently discovered bilayer Ni-based superconductors \cite{Sun2023}. In the context of the latter, our model can be seen as an extension of previously studied mixD $t_\parallel-J_\parallel-J_\perp$ models \cite{Lu2023,Qu2023,Wu2023,Luo2023,Gu2023} to finite-range Coulomb repulsion that should also be present in these materials. Extending our analysis to truely 2D bilayers with $w>2$ at finite doping will be an important step towards a microscopic description of the underlying pairing mechanism in these materials.    \\



\emph{Note added.--} After submission of our manuscript to the pre-print server, we became aware of a closely related work by H.~Yang et al.~\cite{yang2023strong}, in which they use DMRG to study a similar repulsive~$t$-$J$ model on a two-leg ladder. In their work, they also find the emergence of Feshbach resonance and propose a doping induced BEC-to-BCS crossover scenario for the bilayer nickelates.\\

\emph{Acknowledgements.--} We would like to thank Atac Imamoglu, Daniel Jirovec, Felix Palm, Henning Schlömer, Immanuel Bloch, Ivan Morera Navarro, Lieven Vandersypen, Markus Greiner, Matjaz Kebric, Pablo Cova Farina, Tim Harris and Tizian Blatz for helpful discussions. Special thanks to Henning Schlömer for his help with the DMRG implementation of the mixD symmetries. We acknowledge funding by the Deutsche Forschungsgemeinschaft (DFG, German Research Foundation) under Germany's Excellence Strategy -- EXC-2111 -- 390814868 and from the European Research Council (ERC) under the European Union’s Horizon 2020 research and innovation programm (Grant Agreement no 948141) — ERC Starting Grant SimUcQuam. ED acknowledges support from the ARO grant W911NF-20-1-0163 and the SNSF project 200021-212899. HL acknowledges support by the International Max Planck Research School. LH acknowledges support by Studienstiftung des deutschen Volkes.

\bibliography{main.bib}

\bibliographystyle{apsrev4-1}

\newpage~

\appendix
\onecolumngrid
\section*{Supplemental Material}

\subsection{Analysis of low and high doping limits - perturbative analysis}
\label{appendix:low_high_doping}

The black lines in Fig.~\ref{fig:3} are the result of a perturbative expansion for $t_\parallel\ll J_\perp$ in the low and high doping regimes, as in Ref \cite{Bohrdt2021}. In these limits the critical $V_c$ to go from finite binding energies to vanishing binding energies becomes
\begin{align}
    V_c = J_\perp -2t_\parallel+5\frac{t_\parallel^2}{J_\perp}\,
    \label{eq:Vc_low_doping}
\end{align}
for the low doping case and
\begin{align}
    V_c = J_\perp -4t_\parallel+4\frac{t_\parallel^2}{J_\perp}\, 
    \label{eq:Vc_high_doping}
\end{align}
for high doping.

\subsection{The high repulsion limit}
\subsubsection{Derivation of the effective spinon-chargon model}
\begin{figure}[htp]
\centering
\includegraphics[width=0.72\textwidth]{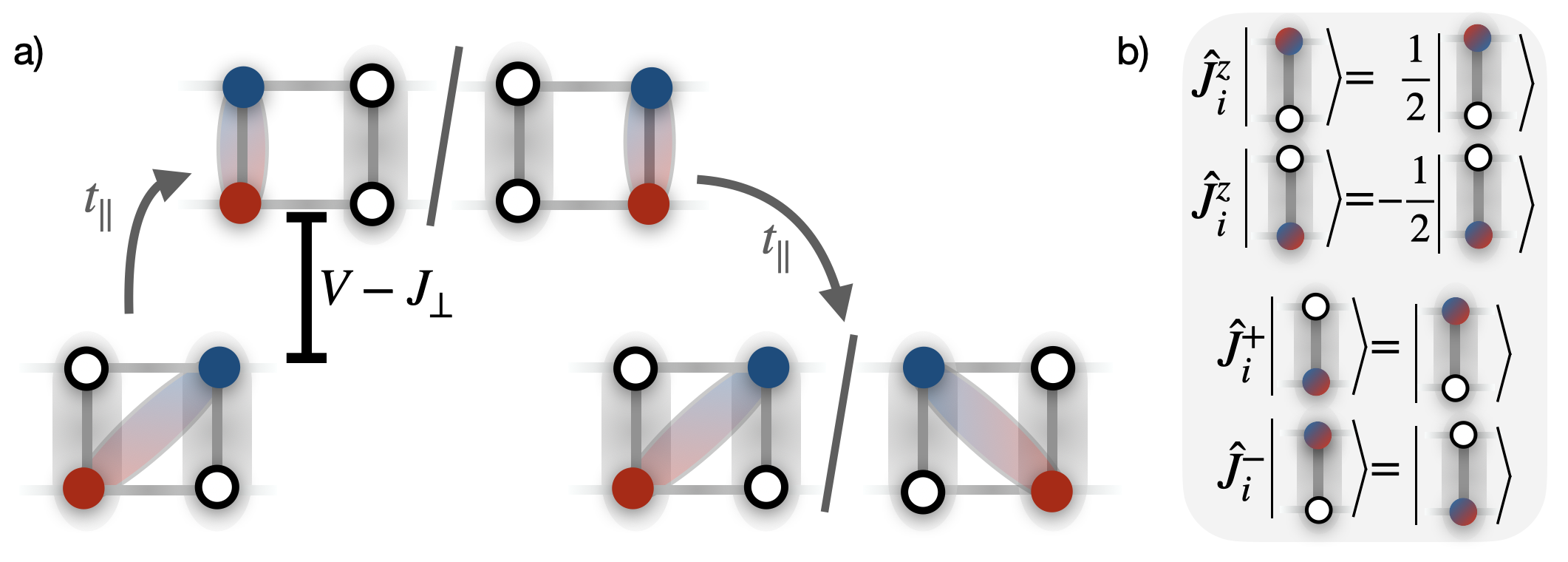}
\vspace{-0.1cm}
\caption{a) Schrieffer-Wolff transformation for $V, V-J_\perp\gg t_\parallel$ with the low-energy sc and high energy cc subspaces. b) Definition of the leg isospin operator $\hat{J}_i$ as introduced formally in Eq.~\eqref{eq:J_operators}. } 
\label{fig:derivation} 
\vspace{-0.4cm}
\end{figure}

The Schrieffer-Wolff transformation that was performed for $V, V-J_\perp\gg t_\parallel$ up to second order in $t_\parallel$ is schematically shown in Fig. \ref{fig:derivation}a. Hereby, the low energy space at strong $V$ is given by sc's, and we consider virtual hopping processes to the high energy subspace given by cc's, separated by a gap $V-J_\perp$. The resulting effective model, Eq. \eqref{eq:H_eff_sc}, involves isospins $\hat{J}$, see Eq. \eqref{eq:J_operators}, that are illustrated Fig. \ref{fig:derivation}b. Further details can be found in Ref. \cite{lange2023feshbach}.

\subsubsection{The bond ordered density wave \label{sec:BODW}}

\begin{figure}[htp]
\centering
\includegraphics[width=0.49\textwidth]{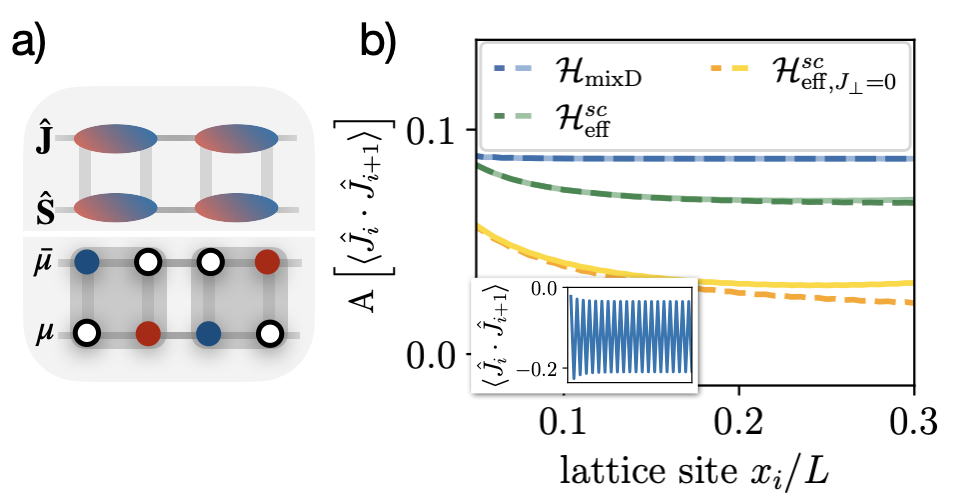}
\caption{a) Illustration of the bond-ordered phase for $\delta= 0.5$ and $V, V-J_\perp\gg t_\parallel$: the system forms plaquettes of $\vec{\hat{S}}$ and $\vec{\hat{J}}$-singlets. b) This yields strong oscillations of $\langle \hat{J}_i\cdot  \hat{J}_{i+1}\rangle$ and $\langle \hat{S}_i\cdot  \hat{S}_{i+1}\rangle$ throughout the system with amplitudes $\mathrm{A} \left[ \langle \hat{J}_i\cdot  \hat{J}_{i+1}\rangle \right]$, which are already visible for moderate $V/J_\perp=5, t_\parallel/J_\perp =1$ (blue lines). The results are compared to predictions by the effective model Eq.~\eqref{eq:Heff_sc_Vlarge_delta0.5} for $J_\perp=1$ (green) and for $J_\perp=0$ (orange) with $L_x=100$ (light solid lines) and $L_x=200$ (dashed). 
} 
\label{fig:4}
\end{figure}

In the large $V$ limit, no more than one sc can occupy each rung. This leads to a charge gap and exponentially decaying pair-pair correlations (see SM \ref{appendix:ppcorr}) at commensurate filling $\delta=50\%$, corresponding to the energy required to create a cc. From our analytical model we hence conclude that the existence of such a charge gap provides a direct signature for the sc-nature of the underlying constituents in the finite-doping regime. 

The remaining spin ($\hat{\vec{S}}_j$) and leg (isospin $\hat{\vec{J}}_j$) degrees of freedom in a ladder at $\delta=50\%$ are described by the effective Hamiltonian
\begin{align} \hat{\mathcal{H}}_\mathrm{eff}^{sc,\delta=\frac{1}{2}}=4\frac{t_\parallel^2}{V} &\sum_{j}\left(\vec{J}_{j+1}\cdot \vec{J}_{j}-\frac{1}{4} \right)\cdot\left[1+\frac{J_\perp}{V}\hat{\mathcal{P}}^S_j\right]    \label{eq:Heff_sc_Vlarge_delta0.5} \end{align}
obtained directly from Eq.~\eqref{eq:H_eff_sc} when $V\gg J_\perp, t_\parallel$ (see \cite{lange2023feshbach}). For large $J_\perp/V$, the ground state of Eq. \eqref{eq:Heff_sc_Vlarge_delta0.5} is a correlated valence-bond crystal (VBS) of both spin and isospins, i.e. an alternating pattern of singlets (no singlets) on bonds $\langle 2j,2j+1 \rangle $ ($\langle 2j+1,2j+2 \rangle $) as illustrated in SM \ref{sec:BODW}, Fig.~\ref{fig:4}a, i.e. it corresponds to a bond-ordered phase of interacting spinon-chargon Cooper pairs (\textit{bond-ordered density wave, BODW}) on plaquettes. This analytical prediction of correlated VBS order in $\hat{\vec{J}}$ and $\hat{\vec{S}}$ is supported even for moderate values of $V/J_\perp=5$ by our numerical results for the mixD model Eq.~\eqref{eq:mixD_Hamiltonian}, see SM \ref{sec:BODW}.
The numerical results at $\delta=0.5$ can be understood in terms of a bond-ordered density wave (BODW). This can be seen from the effective model at half-filling, Eq.~\eqref{eq:Heff_sc_Vlarge_delta0.5}:

At $J_\perp=0$, the ground state of Eq.~\eqref{eq:Heff_sc_Vlarge_delta0.5}
 is a 1D Heisenberg AFM of the isospin $\hat{\vec{J}}$, with power-law correlations, and is fully degenerate in the spin $\hat{\vec{S}}$. In contrast, for large $J_\perp/V$ we expect the ground state to be a correlated valence-bond crystal (VBS) of both spin and isospins, i.e. an alternating pattern of singlets (no singlets) on bonds $\langle 2j,2j+1 \rangle $ ($\langle 2j+1,2j+2 \rangle $) as illustrated in Fig.~\ref{fig:4}a. This state has a lower variational energy contribution for the second term in Eq.~\eqref{eq:Heff_sc_Vlarge_delta0.5} per bond than two independent AFM Heisenberg chains for $\hat{\vec{J}}$ and $\hat{\vec{S}}$ (see \cite{lange2023feshbach}); it corresponds to a bond-ordered phase of interacting spinon-chargon Cooper pairs (\textit{bond-ordered density wave, BODW}) on plaquettes.

 Our analytical prediction of correlated VBS order in $\hat{\vec{J}}$ and $\hat{\vec{S}}$ is supported even for moderate values of $V/J_\perp=5$ by our numerical results for the mixD model, Eq.~\eqref{eq:mixD_Hamiltonian}; blue in Fig.~\ref{fig:4}b. These show that oscillating expectation values $\langle \vec{\hat{J}}_i\cdot  \vec{\hat{J}}_{i+1}\rangle$ and $\langle \vec{\hat{S}}_i\cdot  \vec{\hat{S}}_{i+1}\rangle$ (see inset in Fig.~\ref{fig:4}b), corresponding to the singlet (minima) and no-singlet (maxima) BODW order. 

 While the pure 1D spin-$\frac{1}{2}$ Heisenberg model shows similar VBS-like oscillations in finite-size systems with open boundaries, their amplitude decays notably when increasing the system size. We confirm this behavior for $\langle \hat{\vec{J}}_i\cdot \hat{\vec{J}}_{i+1}\rangle$ at $J_\perp=0$ in Fig.~\ref{fig:4}b (orange), and contrast it with the robust VBS correlations, essentially without any remaining finite size dependence, which we find for $V/J_\perp=5$ at $t_\parallel=J_\perp$, see Fig.~\ref{fig:4}b, both for the mixD model~\eqref{eq:mixD_Hamiltonian} (blue) and the effective sc model~\eqref{eq:H_eff_sc} (green).\\

\subsubsection{Breaking of the SU(2) isospin symmetry}

\begin{figure}[h]
\centering
\includegraphics[width=0.6\textwidth]{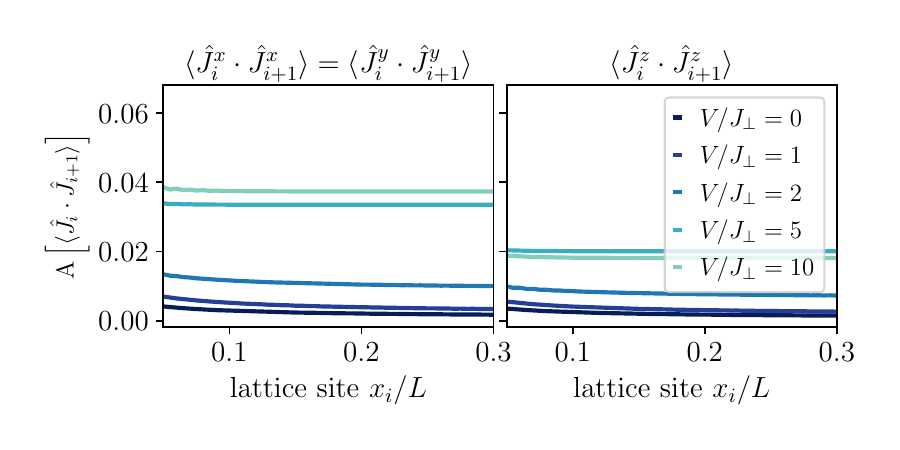}
\caption{Amplitude of the VBS-like oscillations of $\langle \hat{J}_i^a\cdot \hat{J}^a_{i+1}\rangle$ ($a=x,y,z$) for the mixD ladder with $t_\parallel/J_\perp=1$, $V/J_\perp=5$, length $L_x=200$ and $\delta=0.5$.} 
\label{fig:JJ}    
\end{figure}

The effective sc description \eqref{eq:H_eff_sc} is SU(2) symmetric in spins and isospins. However, the SU(2) symmetry of the isospins will break down if higher orders in $t_\parallel/\Delta E$ are considered. This can also be seen in the numerical results when considering the $x$, $y$ and $z$ components of the amplitudes of $\langle \hat{J}_i\cdot  \hat{J}_{i+1}\rangle$ in Fig. \ref{fig:4} separately, see Fig. \ref{fig:JJ}. Note that $\langle \hat{J}_i^x\cdot \hat{J}^x_{i+1}\rangle=\langle \hat{J}_i^y\cdot \hat{J}^y_{i+1}\rangle $  due to total $\hat{J}^z$ conservation. 

\subsection{Pair-pair correlations \label{appendix:ppcorr}}

\begin{figure}[h]
\centering
\includegraphics[width=0.45\textwidth]{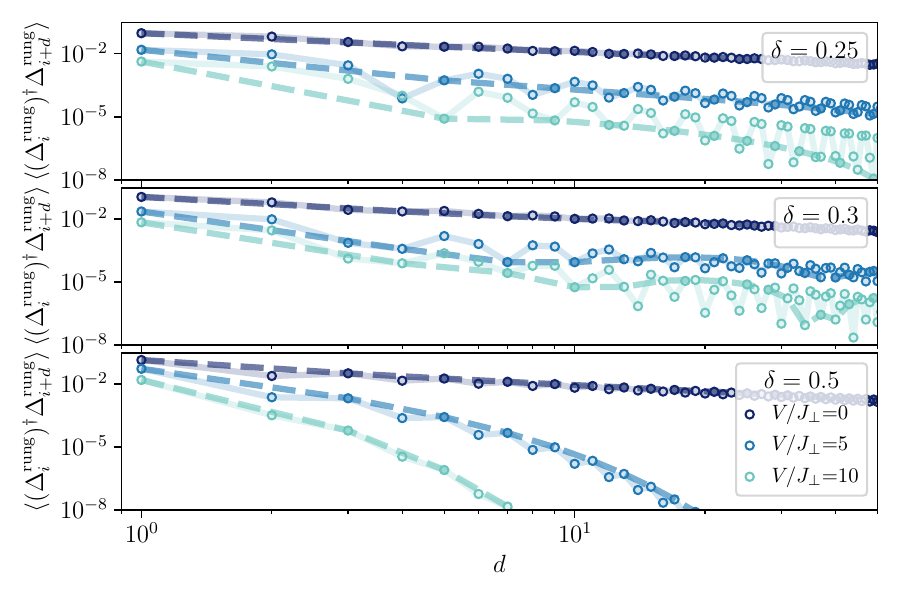}
\includegraphics[width=0.45\textwidth]{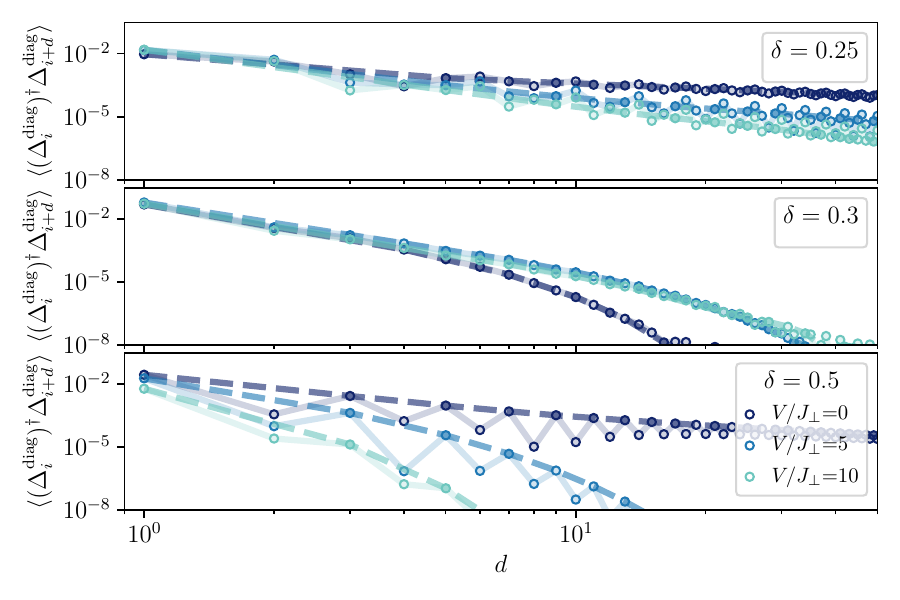}
\caption{Pair-pair correlation $\langle (\Delta_i^{\mathrm{rung}})^\dagger \Delta_{i+d}^{\mathrm{rung}}\rangle $ (left) and $\langle (\Delta_i^{\mathrm{diag}})^\dagger \Delta_{i+d}^{\mathrm{diag}}\rangle $ (right) for $t_\parallel/J_\perp=1$ and different hole dopings, with $\Delta_i^{\mathrm{rung(diag)}}$ defined in Eq. \ref{eq:pair}. The finite doping leads to oscillations of the pair-pair correlations with period $1/\delta$. To guide the eye, we show the correlations for every $1/\delta$ neighbor as dashed lines. }  
\label{fig:pair-pair-corr}    
\end{figure}

Here, we consider the the pair-pair correlations $\langle \Delta_i^\dagger \Delta_{i+d} \rangle $ with the pair annihilation (creation) operator 
\begin{align}
    \Delta_i^{\mathrm{rung}} = \frac{1}{\sqrt{2}}\left(\hat{c}_{i,1,\uparrow}\hat{c}_{i,0,\downarrow}-\hat{c}_{i,1,\downarrow}\hat{c}_{i,0,\uparrow}\right)
    \label{eq:pair}
\end{align}
for pairs with constituents on the same rung, or for pairs extending over neighboring rungs,
\begin{align}
    \Delta_i^{\mathrm{diag}} = \frac{1}{\sqrt{2}}\left(\hat{c}_{i,1,\uparrow}\hat{c}_{i+1,0,\downarrow}-\hat{c}_{i,1,\downarrow}\hat{c}_{i+1,0,\uparrow} \right).
    \label{eq:pair_sc}
\end{align}
The results on both pair-pair correlations are shown in Fig. \ref{fig:pair-pair-corr} for $t_\parallel/J_\perp=1$ and $\delta=0.25, 0.3, 0.5$. The finite doping leads to oscillations of the pair-pair correlations with period $1/\delta$. To guide the eye, we show the correlations for every $1/\delta$ neighbor as dashed lines.

Without repulsion, i.e. for $V=0$, we find algebraically decaying pair-pair correlations for all dopings and both pair-pair correlations $\langle (\Delta_i^{\mathrm{rung}})^\dagger \Delta_{i+d}^{\mathrm{rung}}\rangle $ and $\langle (\Delta_i^{\mathrm{diag}})^\dagger \Delta_{i+d}^{\mathrm{diag}}\rangle $. At large $V/J_\perp$, the overall magnitude of pairs on the same rung decreases and hence also $\langle (\Delta_i^{\mathrm{rung}})^\dagger \Delta_{i+d}^{\mathrm{rung}}\rangle $. However, both pair-pair correlations remain quasi long-ranged with power-law decay, except for $\delta=0.5$, where the charge gap of the BODW leads to exponentially decaying pair-pair correlations.

\subsection{Angle resolved spectroscopy (ARPES) calculations \label{appendix:ARPES}}
The spectral function is given by the Fourier transform of 
\begin{align}
    A_{\vec{i}\vec{j}\sigma}(t) = \langle \psi \vert e^{i\mathcal{\hat{H}}t}\hat{c}^\dagger_{\vec{j}\sigma}e^{-i\mathcal{\hat{H}}t}\hat{c}_{\vec{i}\sigma}\vert \psi\rangle,
\end{align}
where $\psi$ is the ground state, $\hat{c}_{\vec{i}\sigma}^{(\dagger)}$ annihilate (create) perticles at site $\vec{i}$ and $\sigma$ denotes the spin degree of freedom. In order to calculate this quantity, the ground state $\psi$ has to be calculated, $\hat{c}_{\vec{i}\sigma}$ is applied to create a hole, and then the resulting state is time evolved using Krylov and TDVP methods Ref. \cite{PAECKEL2019167998} and linear extrapolation \cite{Barthel2009}. The spectral function is obtained by Fourier transformation w.r.t. spatial indices and time. For the latter, we use $A_\sigma(\vec{k},t)=A_\sigma(\vec{k},-t)^*$ (quasi-steady states) and hence
\begin{align}
    A_\sigma (\vec{k},\omega)= \frac{1}{2\pi}\int_0^{\infty} \mathrm{d}t\, \mathrm{Re}\left[ e^{i\omega t} A_\sigma(\vec{k},t)+ e^{-i\omega t} A_\sigma(\vec{k},-t)\right] = \frac{1}{\pi}\int_0^{\infty} \mathrm{d}t\, \mathrm{Re}\left[ e^{i\omega t} A_\sigma(\vec{k},t)\right].
\end{align}
\end{document}